\documentclass[12pt]{article}
\usepackage{authblk}
\usepackage{fancyhdr}
\usepackage{hyperref}
\usepackage{amsmath, amssymb, amsthm, graphicx, hyperref}
\usepackage[utf8]{inputenc}
\usepackage[a4paper, margin=1in]{geometry}
\usepackage{float}
\usepackage{orcidlink}
\title{AdS Length corrected Thermodynamics of a Black Hole in a Cloud of Strings and Perfect Fluid Dark Matter}
\author[1]{Nazir A. Ganaie \orcidlink{0009-0004-8627-4105} \thanks{nazirahmadgan.82225@jk.gov.in, nazir\_2025phsphy001@nitsri.ac.in}}

\affil[1]{Department of Physics, National Institute of Technology, Srinagar, India-190006}

\date{}

\begin{document}

\maketitle

\begin{abstract}
We investigate the thermodynamic characteristics of the Reisner-Nordstrom black hole in AdS spacetime with a cloud of strinsg and perfect fluid dark matter background, employing a minor correction to the AdS radius. Although the entropy remains unaltered, enthalpy, free energies, pressure, internal energy, and specific heat exhibit considerable modifications. These quantities get enhanced generally by the positive correction  ($\epsilon > 0$) to the AdS scale and lead to faster growth/ slower decay with increase in horizon radius. The opposite effect is reflected by negative corrections ($\epsilon < 0$), which sometimes result in negative or divergent behaviour. We find an interconnection between the combined effect of exotic matter fields and geometric corrections on black hole thermodynamics and thermodynamic stability. 

\end{abstract}

\section{Introduction and Motivation} \label{Introduction}

Black hole thermodynamics offers comprehensive understanding of gravity and spacetime, emerging as a crucial bridge between gravitation, statistical physics and quantum field theory \cite{Bardeen1973, Hooft1993, Seiberg1999}. Recent astrophysical observations have given boost to the field after signatures of dark matter and other exotic energy forms were found in the cosmos. Isolated black holes do not exist; rather they are often embedded in certain matter fields that affect their geometry and thermodynamics, in which the cloud of strings (CoS) and perfect fluid dark matter (PFDM) have been found to be strong candidates.

The CoS model represents a classical limit of string theory, describing string-like matter configurations that can naturally arise in high-energy scenarios \cite{Letelier1979, Rodrigues2018}. 
In contrast, the PFDM model was proposed to explain phenomena on cosmological scale such as galaxy rotation curves \cite{Jusufi2020b, XuWang2018}. Inclusion of such matter sources into the black hole solution affects fundamental properties of the likes of horizon radius, temperature\cite{Hawking1974} and entropy\cite{Bekenstein1973}. 

The interpretation of the cosmological constant $\Lambda$ as the thermodynamic pressure and its conjugate as the thermodynamic volume in the framework of black hole chemistry enriches the thermodynamic study of black holes \cite{Kubiznak2012, Gunasekaran2012}. The AdS black holes undergo Hawking-Page and van der Waals-like phase transitions \cite{Hawking1983, Cadoni2010}. The motivation for our current study to investigate minor corrections to the AdS radius as perturbation comes from the quantum gravity effects, higher order curvature corrections, renormalization of the cosmological constant etc., as these have demonstrated to modify critical conditions and discovery of universal thermodynamic relations \cite{Fernandes2023, PourhassanFaizal2015, Sadeghi2023}, with persepective that minor geometric corrections can affect phase structures, stability and critical phenomena.

Even though black holes have been studied in detail in combined backgrounds of CoS and PFDM, and in various entropy correction schemes, much has not been worked upon AdS length corrected black holes in CoS and PFDM. This study will enhance our understanding of dependence of thermodynamic quantities like enthalpy, specific heat and free energies on the geometric perturbations. 

In this manuscript, we study a spherically symmetric, static RN-AdS black embedded in a combined background of CoS and PFDM \cite{Dharm2025} by incorporating a correction to the AdS radius in the form of a perturbation. We analyze effects on the black hole's entropy, temperature, specific heat and thermodynamic potentials. 

We have organized the mansucript as follows. We present a brief overview of RN-AdS black holes embedded in a CoS and PFDM and evaluate the semi-classical thermodynamic paramters in sec. (\ref{BH with CoS and PFDM}). We derive the corrected thermodynamic parameters, and anlyze thermodynamic stability, as a result of the introduction of the AdS length correction in sec. (\ref{AdS length Corrected Thermodynamics}). And in sec. (\ref{Conclusions}), we present summary of our results and future persepectives.

\section{Black Hole with Cloud of Strings and PFDM} \label{BH with CoS and PFDM}
The solution for RN-AdS black hole embedded in a CoS and PFDM with respective parameters $a$ and $\lambda$, AdS length $l$ of the spacetime, charge $Q$ of the black hole is \cite{Dharm2025}

\begin{align*}
    ds^2 = & -\left(1 - \frac{2M}{r} + \frac{Q^2}{r^2} - a + \frac{\lambda}{r}\ln\left(\frac{r}{\lambda}\right) + \frac{r^2}{l^2}\right) dt^2 \\
    & + \left(1 - \frac{2M}{r} + \frac{Q^2}{r^2} - a + \frac{\lambda}{r}\ln\left(\frac{r}{\lambda}\right) + \frac{r^2}{l^2}\right)^{-1} dr^2 + r^2 d\Omega^2
\end{align*}
 
The largest root of the equation $f(r_+)=0$ at $r=r_+$ gives expression for the black hole mass

\begin{equation}
    M = \frac{r_+}{2}\left(1 + \frac{Q^2}{r_+^2} - a + \frac{r_+^2}{l^2} + \frac{\lambda}{2r_+}\ln\left(\frac{r_+}{\lambda}\right)\right)
\end{equation}
Using $T_H = \frac{f'(r_+)}{4\pi}$, we get expression for the black hole temperature

\begin{equation}
    T_H = \frac{1}{4\pi r_+}\left(1 - a + \frac{3r_+^2}{l^2} - \frac{Q^2}{r_+^2} + \frac{\lambda}{r_+}\right)
    \label{eq:hawking_temp_uncorrected}
\end{equation}
The first law of thermodynamics for the black hole in extended phase space is

\begin{equation}
    dM = T_H\; dS + \Phi\; dQ + \Pi_{\lambda}\; \lambda +\Xi_a\; da+ V\; dP
\end{equation}

where $\Pi_{\lambda}$ denotes the PFDM potential, $\Xi_{a}$ is the CoS potential, and $\Phi$ is the electric potential. 

Assuming constant charge $Q$, the black hole entropy becomes:

\begin{equation}
    S_0 = \int \frac{1}{T_H} \frac{dM}{dr_+} dr_+ = \pi r_+^2 = \frac{A_+}{4}
\end{equation}

Integrating the first law leads to the expression for enthalpy:

\begin{equation}
    H_0 = \int T_H \, dS_0 = \frac{1}{2}\left(\lambda \ln r_+ + \frac{Q^2}{r_+} + (1 - a) r_+ + \frac{r_+^3}{l^2} \right)
\end{equation}

The Helmholtz free energy, obtained via a Legendre transform, is given by:

\begin{equation}
    F_0 = -\int S_0 \, dT_H = \frac{1}{4} \left( 2\lambda \ln r_+ + \frac{3Q^2}{r_+} + (1 - a) r_+ - \frac{r_+^3}{l^2} \right)
\end{equation}

The thermodynamic volume of the black hole is:

\begin{equation}
    V = \frac{4}{3}\pi r_+^3
\end{equation}

Using the relation $P_0 = -\left.\frac{dF_0}{dV}\right|_{T_H}$, the thermodynamic pressure is found to be:

\begin{equation}
    P_0 = -\frac{1 - a - \frac{3Q^2}{r_+^2} + \frac{2\lambda}{r_+} - \frac{3r_+^2}{l^2}}{16\pi r_+^2}
\end{equation}

The internal energy is computed from $U_0 = H_0 - P_0 V$:

\begin{equation}
    U_0 = \frac{1}{12} \left( 2\lambda (1 + 3\ln r_+) + \frac{3Q^2}{r_+} - 7(a - 1) r_+ + \frac{3r_+^3}{l^2} \right)
    \label{eq:internal}
\end{equation}

The Gibbs free energy is determined using $G_0 = F_0 + P_0 V$:

\begin{align}
    G_0 = & -\frac{1}{12} r_+ \left( 1 - a - \frac{3Q^2}{r_+^2} + \frac{2\lambda}{r_+} - \frac{3r_+^2}{l^2} \right) \nonumber \\
    & + \frac{1}{4} \left( 2\lambda \ln r_+ + \frac{3Q^2}{r_+} + (1 - a) r_+ - \frac{r_+^3}{l^2} \right)
    \label{eq:gibbs}
\end{align}

Lastly, the heat capacity $C_0 = T_H \left(\frac{dS}{dT_H}\right)$, which informs us about the thermodynamic stability of the black hole, is given as:

\begin{equation}
    C_0 = \frac{2\pi r_+^2 \left(-l^2 Q^2 + l^2 \lambda r_+ + (1 - a) l^2 r_+^2 + 3r_+^4 \right)}{3l^2 Q^2 - 2l^2 \lambda r_+ + (a - 1) l^2 r_+^2 + 3r_+^4}
\end{equation}

\section{AdS length Corrected Thermodynamics} \label{AdS length Corrected Thermodynamics}

Replacing AdS length by effective AdS length $l \to l/\sqrt{1+\epsilon}$, where $\epsilon \ll 1$ is the correction parameter, we get the BH solution surrounded by CoS and PFDM

\begin{equation} 
\begin{split}
ds^2 = & -\left[1 - a - \frac{2M}{r} + \frac{Q^2}{r^2} + \frac{\lambda}{r} \ln\left(\frac{r}{\lambda}\right) + \frac{r^2}{l^2}(1 + \epsilon) \right] dt^2 \\
& + \left[1 - a - \frac{2M}{r} + \frac{Q^2}{r^2} + \frac{\lambda}{r} \ln\left(\frac{r}{\lambda}\right) + \frac{r^2}{l^2}(1 + \epsilon) \right] dr^2 + r^2 d\Omega^2
\end{split} 
\end{equation}

Using $f(r_+) = 0$, we get the BH mass
\begin{equation}
\label{eq:bh_mass}
M = \frac{(1-a)}{2} r_+ + \frac{Q^2}{2 r_+} + \frac{\lambda}{2} \ln\left(\frac{r_+}{\lambda}\right) + \frac{r_+^3}{2 l^2} (1+\epsilon)
\end{equation}

The equation $\left.\frac{f'(r)}{4\pi}\right|_{r=r_{+}}$ gives the corrected Hawking Temperature
\begin{equation}
\label{eq:hawking_temp}
T_H = \frac{1-a}{4\pi r_+} - \frac{Q^2}{4\pi r_+^3} + \frac{\lambda}{4\pi r_+^2} + \frac{3r_+}{4\pi l^2} (1+\epsilon)
\end{equation}

At constant charge, the entropy of the black hole with correction to AdS length is given by:

\begin{equation}
S_{c} = \int \frac{1}{T_{H}} \frac{dM}{dr_{+}} dr_{+}
\end{equation}

Using (\ref{eq:hawking_temp}) and (\ref{eq:bh_mass}) and integrating, we get the corrected entropy

\begin{equation} \label{eq:cor_entropy}
S_{c} = \pi r_{+}^{2}
\end{equation}

Here we find that the correction to the AdS length does not influence the entropy.

The corrected enthalpy is mathematically given by:

\begin{equation}
H_{c} = \int T_{H} \, dS_{c}
\end{equation}

Evaluating the integral together with (\ref{eq:hawking_temp}) and (\ref{eq:cor_entropy}) gives:

\begin{equation}
H_{c} = \frac{1}{2} \left[ \lambda \ln\left( r_{+} \right) + \frac{Q^{2}}{r_{+}} + r_{+} - a r_{+} + \frac{(1 + \epsilon) r_{+}^{3}}{l^{2}} \right]
\label{eq:enthalpy}
\end{equation}

To observe the effect of a minor correction to the AdS length on the black hole enthalpy, we plot a graph of (\ref{eq:enthalpy}) versus the horizon radius for different values of the correction parameter $\epsilon$.

\begin{figure}[ht]
\centering
\includegraphics[width=0.5\textwidth]{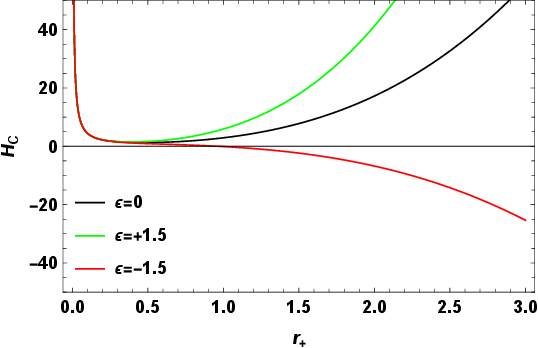}
\caption{Enthalpy $H_c$ v/s horizon radius $r_+$ with different values of $\epsilon$. The curves show the behavior for $\epsilon = 0$ (uncorrected), $\epsilon > 0$, and $\epsilon < 0$.}
\label{fig:enthalpy}
\end{figure}

From the graph (\ref{fig:enthalpy}), we observe that the three curves converge for small black holes, exhibiting positive asymptotic behavior and decrease with horizon radius. While the $-\epsilon$ curve continues decreasing and becomes more negative with increasing black hole size, the $+\epsilon$ curve demonstrates the same trend as the uncorrected one, increasing from $r_{+} \approx 0.5$, growing faster and becoming more positive with increasing black hole size.

The corrected free energy is mathematically given by:

\begin{equation}
F_{c} = - \int S_{c} \, dT_{H}
\end{equation}

Evaluating the integral together with (\ref{eq:hawking_temp}) and (\ref{eq:cor_entropy}) gives:

\begin{equation}
F_{c} = - \frac{1}{4} \left[ 2\lambda \ln\left( r_{+} \right) + \frac{3Q^{2}}{r_{+}} + r_{+} - a r_{+} - \frac{(1 + \epsilon) r_{+}^{3}}{l^{2}} \right]
\label{eq:free_energy}
\end{equation}

To observe the effect of a minor correction to the AdS length on the black hole free energy, we plot a graph of (\ref{eq:free_energy}) versus the horizon radius for different values of the correction parameter $\epsilon$.

\begin{figure}[ht]
\centering
\includegraphics[width=0.5\textwidth]{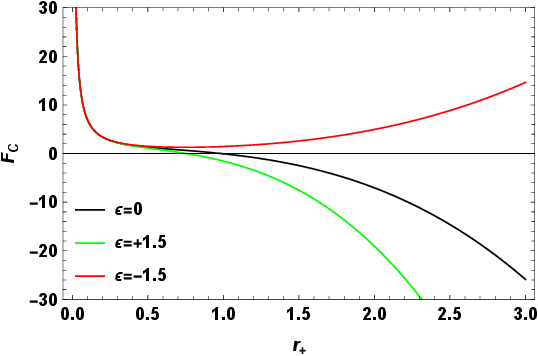}
\caption{Free energy $F_c$ v/s horizon radius $r_+$ with different values of $\epsilon$. The curves show the behavior for $\epsilon = 0$ (uncorrected), $\epsilon > 0$, and $\epsilon < 0$.}
\label{fig:free_energy}
\end{figure}

From the graph (\ref{fig:free_energy}), we observe that the three curves converge for small black holes, exhibiting positive asymptotic behavior and decrease with horizon radius. While the $+\epsilon$ curve demonstrates the same trend as the uncorrected one, continuing to decrease and becoming more negative with increasing black hole size, dropping faster, the $-\epsilon$ curve shows the opposite trend by increasing from $r_{+} \approx 0.5$ and becomes more positive with increasing black hole size.

The volume is given by:

\begin{equation}
V = \frac{4}{3} \pi r_{+}^{3}
\end{equation}

The corrected pressure is given by:

\begin{equation}
P_{c} = - \left. \frac{dF_{c}}{dV} \right|_{T}
\end{equation}

Using the expression for volume together with (\ref{eq:free_energy}), we get:

\begin{equation}
P_{c} = - \frac{1}{16 \pi r_{+}^{2}} \left[ 1 - a - \frac{3Q^{2}}{r_{+}^{2}} + \frac{2\lambda}{r_{+}} - \frac{3(1 + \epsilon) r_{+}^{2}}{l^{2}} \right]
\label{eq:pressure}
\end{equation}

To observe the effect of a minor correction to the AdS length on the black hole pressure, we plot a graph of (\ref{eq:pressure}) versus the horizon radius for different values of the correction parameter $\epsilon$.

\begin{figure}[ht]
\centering
\includegraphics[width=0.5\textwidth]{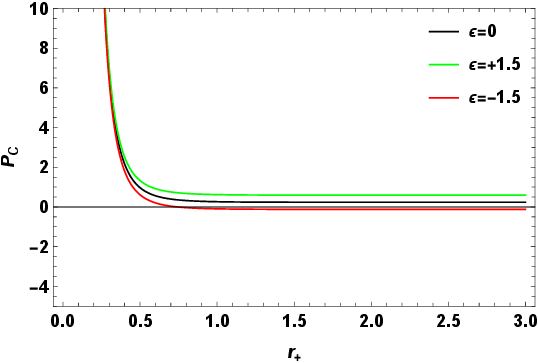}
\caption{Pressure $P_c$ v/s the horizon radius $r_+$ with different values of $\epsilon$. The curves show the behavior for $\epsilon = 0$ (uncorrected), $\epsilon > 0$, and $\epsilon < 0$.}
\label{fig:pressure}
\end{figure}

From the graph (\ref{fig:pressure}), we observe that all three curves show positive asymptotic behavior for small black holes, decreasing with horizon radius and almost attain their respective constant values beyond $r_{+} \approx 0.7$. As the black hole size grows, the positive correction keeps the curve above the uncorrected one, thereby making it drop slower with increasing size, while the negative correction decreases the pressure more and even makes it go into negative values with increasing size.

The corrected internal energy is given by:

\begin{equation}
U_{c} = H_{c} - P_{c} V
\end{equation}

Using (\ref{eq:enthalpy}) and (\ref{eq:pressure}) together with the expression for volume, we get:

\begin{equation}
U_{c} = \frac{1}{12} \left[ 2\lambda (1 + 3 \ln\left( r_{+} \right)) + \frac{3Q^{2}}{r_{+}} - 7(-1 + a) r_{+} + \frac{3(1 + \epsilon) r_{+}^{3}}{l^{2}} \right]
\label{eq:internal_energy}
\end{equation}

To observe the effect of a minor correction to the AdS length on the black hole internal energy, we plot a graph of (\ref{eq:internal_energy}) versus the horizon radius for different values of the correction parameter $\epsilon$.

\begin{figure}[ht]
\centering
\includegraphics[width=0.5\textwidth]{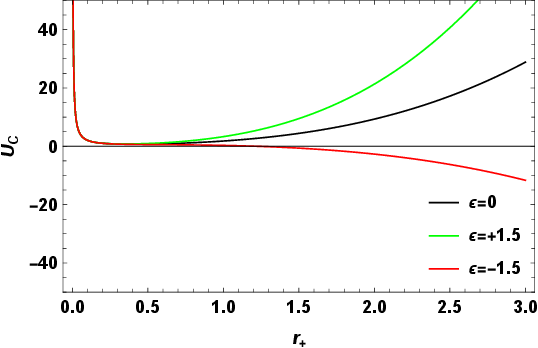}
\caption{Corrected Internal energy $U_c$ v/s horizon radius $r_+$ with different values of $\epsilon$. The curves show the behavior for $\epsilon = 0$ (uncorrected), $\epsilon > 0$, and $\epsilon < 0$.}
\label{fig:internal_energy}
\end{figure}

From the graph (\ref{fig:internal_energy}), we observe that the three curves converge for small black holes, exhibiting positive asymptotic behavior and decrease with horizon radius. While the $-\epsilon$ curve continues decreasing and becomes more negative with increasing black hole size, the $+\epsilon$ curve demonstrates the same trend as the uncorrected one, increasing from $r_{+} \approx 0.5$, growing faster and becoming more positive with increasing black hole size.

The corrected Gibbs free energy is evaluated as:

\begin{equation}
G_{c} = H_{c} - T_{H} S_{c}
\end{equation}

Inserting values, we get:

\begin{equation}
G_{c} = \frac{1}{4} \left[ \lambda (-1 + 2 \ln(r_{+})) + \frac{3Q^{2}}{r_{+}} - (-1 + a) r_{+} - \frac{(1 + \epsilon) r_{+}^{3}}{l^{2}} \right]
\label{eq:gibbs_free_energy}
\end{equation}

To observe the effect of a minor correction to the AdS length on the black hole Gibbs free energy, we plot a graph of (\ref{eq:gibbs_free_energy}) versus the horizon radius for different values of the correction parameter $\epsilon$.

\begin{figure}[ht]
\centering
\includegraphics[width=0.5\textwidth]{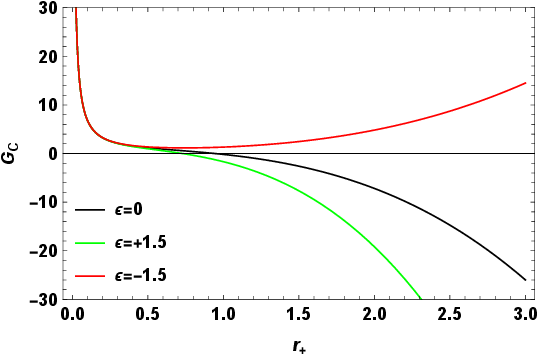}
\caption{Corrected Gibbs free energy $G_c$ v/s horizon radius $r_+$ with different values of $\epsilon$. The curves show the behavior for $\epsilon = 0$ (uncorrected), $\epsilon > 0$, and $\epsilon < 0$.}
\label{fig:gibbs_free_energy}
\end{figure}

From the graph (\ref{fig:gibbs_free_energy}), we observe that the three curves converge for small black holes, exhibiting positive asymptotic behavior and decrease with horizon radius. While the $+\epsilon$ curve demonstrates the same trend as the uncorrected one, continuing to decrease and becoming more negative with increasing black hole size, dropping faster, the $-\epsilon$ curve shows the opposite trend by increasing from $r_{+} \approx 0.5$ and becomes more positive with increasing black hole size.

The expression for the corrected heat capacity is:

\begin{equation}
C_{c} = T_{H} \frac{dS_{c}}{dT_{H}} = T_{H} \frac{\left( dS_{c}/dr_{+} \right)}{\left( dT_{H}/dr_{+} \right)}
\end{equation}

Inserting the values, we get:

\begin{equation}
C_{c} = \frac{2(-l^{2} Q^{2} + l^{2} \lambda r_{+} - (-1 + a) l^{2} r_{+}^{2} + 3 r_{+}^{4}) (-\gamma + \pi \beta r_{+}^{2} + \pi^{2} r_{+}^{4})}{\pi r_{+}^{2} (3 l^{2} Q^{2} - 2 l^{2} \lambda r_{+} + (-1 + a) l^{2} r_{+}^{2} + 3 r_{+}^{4})}
\label{eq:heat_capacity}
\end{equation}

To observe the effect of a minor correction to the AdS length on the black hole heat capacity, we plot a graph of (\ref{eq:heat_capacity}) versus the horizon radius for different values of the correction parameter $\epsilon$.

\begin{figure}[ht]
\centering
\includegraphics[width=0.5\textwidth]{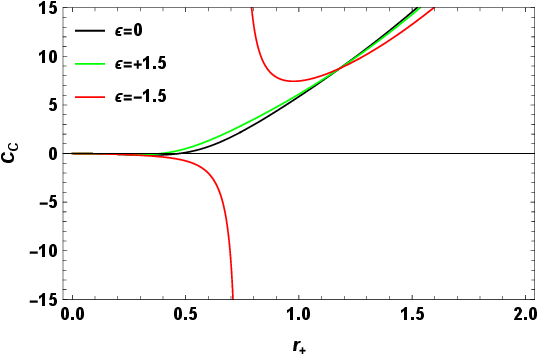}
\caption{Corrected Heat capacity $C_c$ v/s horizon radius $r_+$ with different values of $\epsilon$. The curves show the behavior for $\epsilon = 0$ (uncorrected), $\epsilon > 0$, and $\epsilon < 0$.}
\label{fig:heat_capacity}
\end{figure}

From Fig. (\ref{fig:heat_capacity}), we observe that the three curves converge for small horizon radii, the heat capacity values being negligible, implying instability of small-sized blacks holes, thus substantiating earlier results. The curves depart from one another at $r_{+} \approx 0.3$. The curve corresponding to $+\epsilon$ shows the same trend as the uncorrected curve but increases faster and becomes more positive with increasing black hole size and converges with it at large horizon radii. The negative correction curve exhibits divergence in the heat capacity at the critical point $r_{+} \approx 0.75$, implying charged AdS black holes with CoS and PFDM undergo second-order Davies-type phase transition \cite{Davies1977}. The curve gets closer to the other curves at large size, recovering classical behaviour. 

\section{Conclusions} \label{Conclusions}

In this research paper, we investigated the thermodynamic behavior of a black hole surrounded by a cloud of strings (CoS) and perfect fluid dark matter (PFDM), incorporating a minor correction to the Anti-de Sitter (AdS) length scale in the form of a perturbation. Though the correction did not affect entropy, but other thermodynamic quantities free energies, enthalpy, internal energy, specific heat, pressure, exhibited significant modification. These quantities were found to get enhanced by positive correction, resulting in faster growth/ slowed decay. On the other hand, negative corrections produce the contrasting effect, often giving negative/ divergent behaviour. Our results demonstrate the significance of minor AdS length corrections on thermodynamic behvaiour, stability and critical behaviour of Rn-AdS blacks in exotic matter fields.

\section*{Acknowledgment}
N.A.G. is grateful to wife Mymoona for editing and proof-reading the manuscript.

\end{document}